\documentclass[conference]{IEEEtran}
\IEEEoverridecommandlockouts
\usepackage{array}
\usepackage[hang]{footmisc}

\usepackage[subrefformat=parens,labelformat=parens,caption=false,font=footnotesize]{subfig}
\usepackage{graphicx}
\usepackage[margin = 0.3cm]{caption}
\usepackage{amsmath, amssymb, amsthm}
\usepackage{cite,cleveref}

\usepackage{multirow}

\usepackage[nolist]{acronym}
\begin{acronym}
    \acro{BS}{base station}
    \acro{5G}{fifth generation}
    \acro{HetNets}{heterogeneous networks}
    \acro{CRE}{cell-range expansion}
    \acro{MC}{macro cell}
    \acro{SC}{small cell}
    \acro{QoS}{quality of service}
    \acro{UE}{user equipment}
    \acro{UDN}{ultra dense network}
    \acro{CRE}{cell-range expansion}
    \acro{CMAB}{combinatorial multi-armed bandit}
    \acro{CUCB}{combinatorial upper confidence bound}
    \acro{EE}{energy efficiency}
    \acro{MAB}{multi-armed bandit}
    \acro{RAN}{radio access network}
    \acro{SINR}{signal to interference and noise ratio}
    \acro{PA}{power amplifier}
    \acro{PC}{power consumption}
    \acro{VFA}{value function aprroximation}
    \acro{SARSA}{state-action-reward-state-action}
    \acro{CDSA}{control/data separated architecture}
    \acro{MNO}{mobile network operators}
    \acro{RSRP}{reference signal received power}
    \acro{RL}{reinforcement learning}
\end{acronym}

\usepackage{graphicx,wrapfig}

\newtheorem{remark}{Remark}[]

\crefname{figure}{fig.}{Fig.}
\crefname{section}{sec.}{Sec.}

\usepackage{algorithm}
\usepackage{algpseudocode}
\usepackage{scalerel}
\usepackage{multicol}

\usepackage{textcomp}

\usepackage{bbm}
\usepackage{xcolor}
\usepackage{dsfont}
\usepackage{bm}

\usepackage{caption}
\usepackage{mathtools}

\def\BibTeX{{\rm B\kern-.05em{\sc i\kern-.025em b}\kern-.08em
    T\kern-.1667em\lower.7ex\hbox{E}\kern-.125emX}}

\usepackage{titlesec}
\titlespacing{\section}{1pt}{1pt}{1pt} 
\begin{document}

\title{\fontsize{22.8}{27.6}\selectfont Joint Sleep Mode Activation and Load Balancing with Dynamic Cell Load: A Combinatorial Bandit Approach}


\author{
    Wajahat Bashir Gilkar and 
    Gourab Ghatak,~\IEEEmembership{Member ~IEEE} \\
    Department of Electrical Engineering,
    Indian Institute of Technology Delhi, New Delhi-110016, India \\
    Email: eez238361@iitd.ac.in, gghatak@ee.iitd.ac.in
}

\maketitle

\begin{abstract}
We propose a combinatorial bandit formulation to opportunistically trigger sleep modes in gNode-B (gNB) small cells (SCs), followed by a cell range expansion (CRE)-based load balancing procedure. This is implemented by ensuring that the fifth generation (5G) quality of service identifier (5QI)\mbox{-requirements} of user equipments (UEs) are maintained. The key challenge is the fact that while deactivating a given SC gNB reduces its own consumption, it may increase the load on neighboring gNBs and the macro gNB (coverage cell), impacting the overall energy efficiency. This phenomenon is accurately characterized by modeling the dynamic cell load that jointly takes into account the location of the UEs, their relative locations to all the SCs, and their data demands. We experimentally show that the proposed combinatorial upper confidence bound (CUCB) followed by the load balancer outperforms not only the naive strategies like arbitrarily keeping all the SCs on, but also other state-of-the-art reinforcement learning solutions. The proposed algorithm can be implemented as open-radio access network (O-RAN) near-real-time (NRT) RAN intelligent controller (RIC) xApps.
\end{abstract}


\section{Introduction}
Ultra-dense deployment of new-radio (NR) \acp{SC} in \ac{5G} \ac{HetNets}, as standardized in 3GPP, is a common strategy adopted by mobile network operators to reduce path-loss and improve cell-edge \ac{UE} throughput, while supporting ITU-IMT 2020 requirements for capacity and energy efficiency~\cite{kooshki2022energy}. However, with an increase in the \ac{SC} density, the number of network components also increases leading to higher \ac{PC} and emissions. Naturally, \acp{MNO} are concerned about the high \ac{PC} and are investigating solutions to solve it through new network planning and optimization approaches with \ac{EE} at its core~\cite{kolta2021going}. It is important to note that \acp{BS} consume $60\%$ to $80\%$ of the total network's energy~\cite{feng2017base}, making them the most power-hungry elements of the \ac{RAN}. Optimizing their \ac{PC} may result in a significant reduction in the \ac{PC} of the network and enhance its \ac{EE}.

In literature, several techniques have been proposed to reduce the \ac{PC}, among which, \ac{SC} sleep mode strategy is particularly attractive as it does not require changes in the network architecture~\cite{wu2015energy}. An efficient implementation of \ac{SC} sleep strategy necessitates the consideration of the spatio-temporal network load due to the inherent trade-off between the \ac{QoS} of the \acp{UE} and the network's energy savings. Set in this context, we consider a \ac{CDSA}~\cite{mohamed2015control}, where \acp{MC} act as control \acp{BS} to ensure continuous coverage while handling signaling functions and low data rate services. On the other hand, \acp{SC} serve as capacity cells to provide high data rate services and are connected to \acp{MC} through an appropriate backhaul. We then propose a \ac{CMAB} model to formalize the cell deactivation/activation decisions for gNB-SCs (aligned with 3GPP energy saving management functions in TS 28.310), jointly with load-balancing procedures (via \ac{CRE} optimization) that can be implemented as open-\ac{RAN} near-real-time (NRT) RAN intelligent controller (RIC) xApps. As a solution, we employ the \ac{CUCB} algorithm to determine the activity state of the \acp{SC}.

{\bf Related work:}
The authors of \cite{sun2018energy} have studied the problem of \ac{UE} association and \ac{SC} sleeping in multi-tier dense networks posed as a complex mixed integer programming. They have proposed two low-complexity heuristic algorithms that determine the optimal \ac{UE} association and the \ac{SC} switching pattern. In \cite{wu2016dynamic},
a greedy heuristic algorithm is used to find the optimal \ac{SC} ON/OFF configuration. Heuristics in general computationally expensive to implement, hence have limited applicability.
On the contrary, \ac{RL} approaches have gained traction for this problem, e.g., in \cite{abubakar2019q}, a two-tier HetNet with optimal switching and traffic offloading strategy was determined by leveraging a Q-learning framework. Similarly, Salem et al. in \cite{salem2018reinforcement} have employed a Q-learning algorithm to determine the amount of time that the \ac{SC} should spend in a particular sleep level so that there is a fair trade-off between the energy savings and delay.
Ozturk {\it et al.}~\cite{ozturk2021energy} have studied the problem of minimizing \ac{PC} using SARSA with value function approximation while
guaranteeing \acp{UE}' \ac{QoS}. From a \ac{MAB} perspective, \cite{ayala2019online} investigated joint energy savings and interference coordination and proposed a two-level algorithm to determine the most energy-efficient control actions. The key challenge in employing classical \ac{MAB} techniques to this problem is indeed the combinatorial nature of the action set, which grows detrimentally with deployment density.

{\bf Research gap and contributions:} None of the above works consider the dynamic cell load of the \acp{SC} and the \ac{MC} for modeling \ac{PC}. To elaborate, the authors in \cite{ghatak2018accurate} showed that the spatio-temporal load of a cell depends not only on the number of the connected \acp{UE} and their corresponding data demands, but also on their relative locations with respect to the serving and interfering \acp{BS}. Thus, the impact of turning off a given \ac{SC} on the overall network load depends on the locations of the connected \acp{UE} to the \ac{SC} with respect to the other \acp{SC} and the \ac{MC} to which they are potentially offloaded. Ours is the first work that takes into account such dynamic cell load in order to optimize the sleep mode decisions.
Overall, compared to the previous literature, the contributions of the paper are as follows:
\begin{itemize}
    \item The SC sleeping strategy is modeled as a combinatorial bandit problem with super arm rewards dependent on dynamic cell load, leading to a non-monotonic reward function--an issue that has not yet been addressed in literature.
    For this problem, a \ac{CMAB} algorithm optimized for monotonic rewards is proposed to find the best sleep configuration. The results confirm the efficacy of the algorithm even in this non-monotonic setting--for which no provable solution exists.
    
    \item Once the optimal ASM configuration is obtained, it can be applied via O-RAN near-RT RIC xApps interfacing with the E2 node of gNB-SCs, and remains valid for several traffic intervals. To further improve \ac{EE}, a load-balancing module is proposed, which can be mapped to the self-organizing network (SON) functions defined in 3GPP TS 32.500~\cite{jorguseski2014self}. This is done by optimizing the \acp{CRE} parameters of the \acp{SC} that are kept on.
    \item We compare our solution with the \texttt{ALL-ON} strategy and \texttt{VFA-SARSA} based strategy proposed in \cite{ozturk2021energy} and demonstrate that the proposed solution outperforms both of them. 
    The reason for this observation is the consideration of dynamic cell load and \ac{CRE} optimization.
    
\end{itemize}

\section{Dynamic Cell Load and Power Consumption}
{\bf Network geometry and propagation:} Consider an ultra-dense \ac{RAN} where $L$ number of co-channel \acp{SC} are deployed under the coverage area of one \ac{MC}. In the 5G and beyond architecture, the \ac{MC} acts as the coverage cell with opportunistic data services while the \acp{SC} are the capacity cells. The O-RAN near-RT RIC, via its E2 interface, dynamically configures the \ac{CRE} offsets of \ac{SC} gNBs to optimize \ac{UE} association and execute load balancing, in line with 3GPP self-organizing network (SON) and radio-resource management (RRM) functions in TS 32.500~\cite{jorguseski2014self}. Consider $\phi_i$ to be the \ac{CRE} parameter of the $i-th$ \ac{SC} and $\phi_M$ to be the \ac{CRE} parameter of the \ac{MC}. Let the transmit powers of \ac{SC} and \ac{MC} be $P_S$ and $P_M$ respectively. Additionally consider the 3GPP path-loss model with path loss exponent $\alpha_p$ and path-loss constants $K_S$ and $K_M$, for the \acp{SC} and the \ac{MC}, respectively. Furthermore, let us denote the relative distances between the \ac{UE} and the \ac{SC} $i$ as $d_{ui}$, while the distance between the \ac{UE} and the \ac{MC} is $d_{uM}$. The \ac{UE} association is based on the \ac{RSRP}, thus a \ac{UE} $u$ gets connected to the $i$-th SC if:
\[
\phi_i P_S K_S d_{ui}^{-\alpha_{p}} > \max_{l \neq i}\{\phi_l P_S K_S d_{ul}^{-\alpha_{p}}, \phi_M P_M K_M d_{uM}^{-\alpha_{p}} \}.
\]
The \ac{UE} $u$ gets associated to the \ac{MC} if it receives the highest received power from it i.e.,
\[\phi_M P_M K_M d_{uM}^{-\alpha_{p}} > \phi_i P_S K_S d_{ui}^{-\alpha_{p}}, \; \forall \; i \in \{1,2,....,L\}.
\]
The \ac{SINR} experienced by the \(j\)-th \ac{UE} connected to the \ac{MC} is
\begin{align}
    \chi_{jM} = \frac{P_M K_M d_{jM}^{-\alpha}}{N_0 + \sum_{l=1}^L P_S K_S d_{jMl}^{-\alpha} \delta_l }
\end{align}
where the second term in the denominator represents the interference experienced by the \ac{UE} from the \acp{SC} in the network. $d_{jMl}$ is the distance of \(j\)-th \ac{UE} connected to the \ac{MC} from \(l\)-th \ac{SC}, $N_0$ represents the noise power and $\delta_l \in \{0,1\}$ represents the state of the $l-th$ \ac{SC} i.e., whether ON or OFF. Similarly, the \ac{SINR} experienced by the $j$-th \ac{UE} associated with the $i$-th \ac{SC} is
\begin{align}
    \chi_{ji} = \frac{P_S K_S d_{j,i}^{-\alpha}}{N_0 + P_M K_M d_{jiM}^{-\alpha}+ \sum_{l=1,l \neq i}^L P_S K_S d{jil}^{-\alpha} \delta_l}
\end{align}
where the second and third terms in the denominator represent the interference experienced by the \ac{UE} from the \ac{MC} and other \acp{SC} in the network, respectively.

{\bf Traffic model:} Let there be $N$ total \acp{UE} in the network, where $N = N_M + \sum^L_{i=1}N_i$. Here, $N_M$ and $N_i$ are the number of \acp{UE} served by the \ac{MC} and the \(i\)-{th} \ac{SC}, respectively. Additionally, let us denote by $B_M$ and $B_S$ the bandwidth of the \ac{MC} and each \ac{SC}, respectively. The \ac{UE} requests arrive as a Poisson arrival process with an average arrival rate of $\lambda_a$ /s/m$^2$. Each user-request consists of a data-size of $\sigma$ bits, resulting in a traffic of $\omega = \sigma\lambda_a$ bits/s/m$^2$. The downlink data rate experienced by a \ac{UE} connected to the \ac{MC} is thus
$R_{jM} = \frac{B_M }{N_M} \log(1 + \chi_{jM}).$
Consequently, the load generated by this \ac{UE} on \ac{MC} is 
\begin{align}
\rho_{jM} = \frac{\omega}{R_{jM}} 
= \frac{\omega N_M}{B_M \cdot \log(1 + \chi_{jM})}
\end{align}
As a result the total load on \ac{MC} due to all the $N_M$ \acp{UE} associated with it is
\begin{align}
\rho_M = \sum_{j=1}^{N_M} \rho_{jM} = \sum_{j=1}^{N_M}\frac{\omega \cdot N_M}{B_M \cdot \log(1 + \chi_{jM})}.
\end{align}
\begin{remark}
    It is important to highlight that this formulation of the cell load is distinct from other works that study energy management in HetNets. In particular, this formulation not only accounts for the number of connected \acp{UE} and their data-requests, but more importantly, their locations and experienced \ac{SINR}. This is a more accurate description of the actual dynamic cell load.
\end{remark}

Similarly, the data rate of the $j$-th \ac{UE} associated to the $i$-th cell is $R_{ji} = \frac{B_S}{N_i}\cdot \log(1 + \chi_{j,i})$, thus generating a corresponding cell load $\rho_{j,i} = \frac{\omega}{R_{j,i}} = \frac{\omega \cdot N_i}{B_S \cdot \log(1 + \chi_{ji})}$.
Thus, the total load on $i$-th SC is
\begin{align}
\rho_{i} = \sum_{j=1}^{N_i} \rho_{ji} = \sum_{j=1}^{N_{i}} \frac{\omega \cdot N_i}{B_S \cdot \log(1 + \chi_{ji})}.
\end{align}

{\bf Power consumption model}
Following the energy-aware radio and network technologies (EARTH) power consumption model~\cite{auer2011much} , the instantaneous \ac{PC} $\xi_i$ of $i-th$ \ac{SC} is 
\begin{align}
\xi_i = 
\begin{cases} 
P_{oi} + \eta_i \rho_i P_S, & \delta_i = 1, \\
P_{\rm sleep}, & \delta_i = 0
\end{cases}
\end{align}
where $P_{oi}$ and $P_{\rm sleep}$ are the operational and sleep circuit \ac{PC} of \(i\)-th \ac{SC}, respectively, and \(\eta_i\) is the \ac{PA} efficiency. The \ac{MC} is always in the ON state as it is the coverage cell. The \ac{PC} of the \ac{MC} is
$\xi_M = P_{oM} + \eta_M \rho_M P_M$,
where $P_{oM}$, $\eta_M$, $\rho_M$ and $P_M$ are the operational circuit \ac{PC}, \ac{PA} efficiency, load factor and transmit power of the \ac{MC}, respectively. The instantaneous \ac{PC} $\xi_T$ of the network is
\begin{align}
\xi_T = \xi_M + \sum_{i=1}^L \xi_i.
\end{align}

\section{Problem Formulation and \ac{CMAB} Framework}
The objective is to find a sleep configuration $\{\delta_1, \delta_2, \dots, \delta_{L}\} \in \{0,1\}^L$ that minimizes the power consumption of the network such that all the \acp{UE} in the network experience a guaranteed bit-rate $R_{min}$. Note that each time the state of a \ac{SC} is changed, say from 1 to 0, the connected \acp{UE} to this \ac{SC} must be preemptively offloaded to other \acp{SC} active in the network or the \ac{MC} based on next-best RSSI measurements. This not only changes the load of the new cell but also reconfigures the overall network energy consumption. Furthermore, it alters the experienced \ac{SINR} experienced by the transferred \acp{UE}. The same occurs in case when the configuration of a \ac{SC} changes from 0 to 1. We state the target optimization problem as
\begin{align*}
    \min_{\{0,1\}^L} \quad &\xi_T \\ 
    \text{s.t.} \quad &\rho_M \leq 1; \;\;\rho_i \leq 1, \;\; \forall  i \in \{1,2,....,L\}   \\
    &R_{jM} \geq R_{\rm min}; \;\; R_{ji} \geq R_{\rm min}, \;\; \forall i \in \{1,2,....,L\}  
\end{align*}
The first constraint ensures that none of the cells in the network are overloaded. The last constraint ensures the delivery of the minimum guaranteed bit-rate to the \acp{UE}.

To solve the above we reconstruct the problem in the \ac{CMAB} framework, since classical online learning or bandit models will not scale with the number of \acp{SC}. For this, let the state of each \ac{SC} represent an arm (here, simple arm). In \ac{CMAB}, the player does not select a simple arm but a subset of simple arms called a {\it super-arm}, which provides a super-arm reward to the player. The super-arm reward in the classical \ac{CMAB} problem depends on the rewards of the constituent simple arms. As a super-arm is played, rewards of simple arms that are a part of the super-arm are also revealed, which are used in the selection of super-arms in future rounds. Then, the goal is to maximize the cumulative super-arm reward over the course of time.
\vspace{-0.5em}
\begin{remark}
    Contrary to the classical \ac{CMAB} setting, in our problem, the reward on playing a super-arm not only depends on the constituent simple arms but also a penalty term due to the detrimental impact on the other arms. This further implies that the regret guarantees due to the monotonicity assumption in existing \ac{CMAB} algorithms (e.g., see~\cite{chen2013combinatorial}) may not hold.
\end{remark}
\vspace{-0.5em}
Let us mathematically formulate the \ac{CMAB} problem. Let simple arms correspond to setting $\delta_l = 0$ for an \ac{SC} $l$. In addition, the player has the option to not play any arm in a time step, i.e., $\delta_l = 1, \forall l\in [L]$. At each time-slot the mean reward of the simple arm $\mu_l$ is defined as
\begin{align}
    \mu_l = \mathbb{E}\left[P_{l,\rm all} - P_{\rm sleep}\right],
\end{align}
where $P_{i,\rm all}$ is the power consumed by the $l$-th \ac{SC} when all the \acp{SC} are turned on. The expectation is with respect to the randomness in $\omega$.
\vspace{-0.5em}
\begin{remark}
The above formulation for the mean reward compares the performance of a simple action from the perspective of a static decision -- that of keeping all \acp{SC} on. This is different from the instantaneous gain in power savings, i.e., savings as compared to the power consumption just before the decision. This formalization makes the problem environment stateless and stationary, so as to allow the application of \ac{CMAB} algorithms.
\end{remark}
\vspace{-0.5em}
On playing a super-arm, define $\Delta{P_M}=\eta_M P_M(\rho_{M,\rm all}- \hat \rho_M)$ as the power change for the \ac{MC} and $\Delta{P_i}=\eta_i P_i (\rho_{i,\rm all}-\hat \rho_i)$ as power change for the $i-th$ \ac{SC}, where $\hat \rho_M$ and $\hat\rho_i$ are the estimated load values of \ac{MC} and $i-th$ \ac{SC} when a super-arm is played, and $\rho_{M,\rm all}$ and $\rho_{i,\rm all}$ are load values of \ac{MC} and $i-th$ \ac{SC} if the \texttt{ALL-ON} strategy was implemented on the network.
Based on this, we define the expected reward of super-arm $S \subseteq [K]$ as
\begin{algorithm}[t]
\caption{\(\alpha,\beta\)-Approximation Oracle}
    \begin{algorithmic}[1]
        \State \textbf{Input:} The mean vector \({\bf\bar\mu} = \{\bar\mu_1, \bar\mu_2,....,\bar\mu_m \}\) of simple arm rewards, \(\alpha \in [0,1] \)
        \For{each super-arm \(S \in \mathcal{S}\)}
             \State Compute \(r_{\bf \bar\mu}(S) \) using the vetor \(\bar\mu\).
        \EndFor
        \State Compute the optimal super-arm reward with respect to the given mean vector \(\bf\bar\mu\):\\
        \(r_{\bf \bar\mu}(S^*) = \max_{S \in \mathcal{S}} r_{\bf \bar\mu}(S)\)
        \State Determine the \(\alpha-\)optimal super-arms set as:\\
        \(\mathcal{S'} = \{S \in \mathcal{S}: r_{\bf \bar\mu}(S) \geq \alpha.r_{\bf \bar\mu}(S^*) \}\)

        \State With probability \(\beta\), return the super-arm \(S \in \mathcal{S'}\).             
    \end{algorithmic}
\end{algorithm}
\begin{align}
      r_{\bf \mu}(S) = &\underbrace{\sum_{i=1}^L \mu_i(1-\delta_i)}_{A} + \underbrace{\Delta{P_M}}_{B} + \underbrace{\sum_{i=1}^L \Delta{P_i} \delta_i}_{C} \\ 
      &- \underbrace{\mathcal{P}_1 \max(0, (\hat\rho_M -1)) - \sum_{i=1}^L \mathcal{P}_2 \max(0, (\hat \rho_i -1))}_{D} \nonumber
\end{align}

The term $A$ accounts for the power savings in the \acp{SC} that are turned off. The term $B$ is the increase/decrease in the \ac{PC} of the \ac{MC} due to turning off some \acp{SC}. The term $C$ accounts for the increase/decrease in the \ac{PC} of the \acp{SC} that are kept on. Finally, the term $D$ accounts for violations in the overload criterion, where $\mathcal{P}_1$ and $\mathcal{P}_2$ are the penalty terms.

\begin{algorithm}[b]
\caption{CUCB with Computation Oracle}
\begin{algorithmic}[1]

    \State \textbf{For each simple arm} \(a_i\) Maintain:
    \State (1) variable \( T_i \) as the total number of times arm \( a_i \) is played so far.
    \State (2) variable \( \hat{\mu}_i \) as the estimated mean of all outcomes \( r_{i} \)'s of arm \( i \) observed so far.
    
    \For{each arm \( a_i \)}
        \State Play an arbitrary super-arm \( S \in \mathcal{S} \) such that \( i \in S \).
        \State Update variables \( T_i \) and \( \hat{\mu}_i \).
    \EndFor
    
    \State \( t \gets m \).
    
    \While{$t \leq n$}
        \State \( t \gets t + 1 \).
        \For{each arm \( i \)}
            \State Set \( \bar{\mu}_i = \hat{\mu}_i + \sqrt{\frac{3 \ln t}{2 T_i}} \).
        \EndFor
        \State \( S = \text{Oracle}(\bar{\mu}_1, \bar{\mu}_2, \dots, \bar{\mu}_m) \).
        \State Play \( S \) and update all \( T_i \)'s and \( \hat{\mu}_i \)'s.
    \EndWhile
\end{algorithmic}
\end{algorithm}
\vspace{-0.5ex}
\subsection{$(\alpha,\beta)$-Approximation Oracle and \ac{CUCB}}
Recall that due to the NP-hard nature of this combinatorial problem, most \ac{CMAB} algorithms rely on an $\alpha,\beta$-approximation oracle for action selection. In particular, for some $\alpha,\beta \leq 1$, the algorithm outputs a super-arm $S$ such that $\mathbb{P}\left(r_{\mu}(S) \geq \alpha.r_{\bf \mu}(S^*)\right) \geq \beta$, where  $r_{\bf \mu}(S^*)= \max_{S \in \mathcal{S}} r_{\bf \mu}(S)$ is the optimal super-arm reward as shown in algorithm 1. With this formulation, since only a $\beta$ fraction of the oracle calculations are successful, and when successful, the reward is only $\alpha$-approximate of the optimal value, the performance is benchmarked with respect to the $\alpha,\beta$-approximation regret defined
\begin{align}
    {\rm Reg}_{\mu,\alpha,\beta}(n) = n\alpha\beta r_{\bf \mu}(S^*) - E_{S}\left[\sum^n_{t=1} r_{\mu}(S_t )\right].
\end{align}
Now we present the \ac{CUCB} algorithm shown in algorithm 2. After first $m$ number of initialization rounds (equal to the number of simple arms, which here is the number of \acp{SC}), we maintain an empirical mean $\hat\mu_i$ for each arm. The expectation vector $\hat{\bf \mu}$ is fed to the $(\alpha,\beta)$-approximation oracle with an added confidence interval (that depends on the number of times the corresponding arm has been played, $T_i$) together forming the upper confidence bound. The oracle then returns an $\alpha$-optimal super-arm with respect to the mean vector $\bar\mu$, which is then played, and the variables \(T_i\)'s and \(\hat\mu_i\)'s are updated accordingly.

Note that due to the non-monotonic nature of the super arm reward, the $(\alpha,\beta)$-approximation oracle exhaustively evaluates the reward of each super arm. However, the expected base arm rewards required to compute these super arm rewards are still learned by exploiting the structured framework of \ac{CMAB}, which leads to fast learning. 

\subsection{CRE Optimization}
A direct application of the super-arm $S$ returned by the $(\alpha, \beta)$-approximation oracle may not be much energy efficient as it may increase the load of the \ac{MC} and the \acp{SC} that are kept on due to inefficient load balancing. Therefore, to make more energy savings, we propose optimizing the \ac{CRE} parameters of the \acp{SC} that are kept on according to the super-arm $S$. The \ac{CRE} optimization is formulated as
\begin{align}
    \min_{\Phi} \rho_M + \sum_{i \in S} \rho_i \delta_i \hspace{2.cm} \nonumber \\
    \text{s.t} \quad \rho_M \leq 1 ; \;
    \rho_i \leq 1, \forall \hspace{0.2cm} i \in S.  \nonumber
\end{align}

\begin{algorithm}[t]
\caption{Powell's Optimization Method}
\begin{algorithmic}[1]
\State \textbf{Input:} Initial point $\mathbf{\Phi}_0$, function $f$, tolerance $\epsilon_p$, maximum iterations $N$
\State \textbf{Initialize:} Set direction vectors $\{\mathbf{d}_1, \mathbf{d}_2, \dots, \mathbf{d}_n\}$ as unit vectors
\For{$k = 1$ to $N$}
    \State $\mathbf{\Phi}_{start} \gets \mathbf{\Phi}_0$
    \For{$i = 1$ to $n$}
        \State Perform line search: find $\alpha_i$ minimizing $f(\mathbf{\Phi}_{i-1} + \alpha_i \mathbf{d}_i)$
        \State Update: $\mathbf{\Phi}_i \gets \mathbf{\Phi}_{i-1} + \alpha_i \mathbf{d}_i$
    \EndFor
    \State $\mathbf{d}_{n+1} \gets \mathbf{\Phi}_n - \mathbf{\Phi}_0$
    \State Perform line search along $\mathbf{d}_{n+1}$: find $\alpha$ minimizing $f(\mathbf{\Phi}_n + \alpha \mathbf{d}_{n+1})$
    \State Update: $\mathbf{\Phi}_{new} \gets \mathbf{\Phi}_n + \alpha \mathbf{d}_{n+1}$
    \If{$\|\mathbf{\Phi}_{new} - \mathbf{\Phi}_0\| < \epsilon_p$}
        \State \textbf{Break}
    \EndIf
    \State Update direction set: replace one $\mathbf{d}_i$ with $\mathbf{d}_{n+1}$
    \State $\mathbf{\Phi}_0 \gets \mathbf{\Phi}_{new}$
\EndFor
\State \textbf{Output:} Approximate minimum point $\mathbf{x}_{new}$
\end{algorithmic}
\end{algorithm}

First we set $\phi_i = 0, \forall i \in S$. Then, we note that computing the derivative of the objective function with respect to $\Phi$ is infeasible. To alleviate this issue, we employ a derivative-free Powell's conjugate direction method (shown in Algorithm 3) for finding a local minimum of a function $f: \mathbb{R}^n \rightarrow \mathbb{R}$ without using it's derivatives~\cite{powell1964efficient}. The algorithm performs successive one‑dimensional minimization (using line search) along a set of $n$ initial, mutually orthogonal directions. It is initialized with an initial point $\Phi_0$ and a set of search direction vectors $\{d_1, d_2,...,d_n\}$. It, starting from $\Phi_0$, minimizes $f$ sequentially along each direction $d_i$, updating the current point after each line search. After cycling through all $n$ directions, a new direction equal to the net displacement from the starting point is defined and an additional line search is performed along it; if the change in this minimization is very small (less than $\epsilon_p$ ), the algorithm stops. Otherwise, it replaces one of the old directions with the new displacement direction and repeats the process up to a maximum of $N$ iterations, thereby adaptively building approximate conjugate directions and converging efficiently.

\begin{figure*}[h]
\centering
\subfloat[]
{\includegraphics[width=0.32\textwidth]{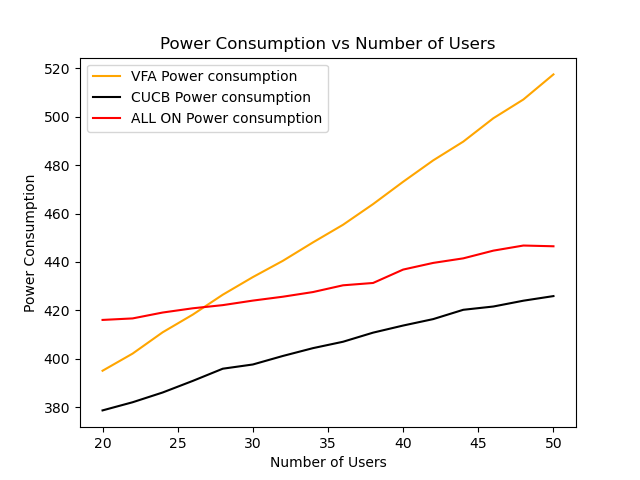}
\label{fig: PC_vs_UE_Num}}
\hfil
\subfloat[]
{\includegraphics[width=0.32\textwidth]{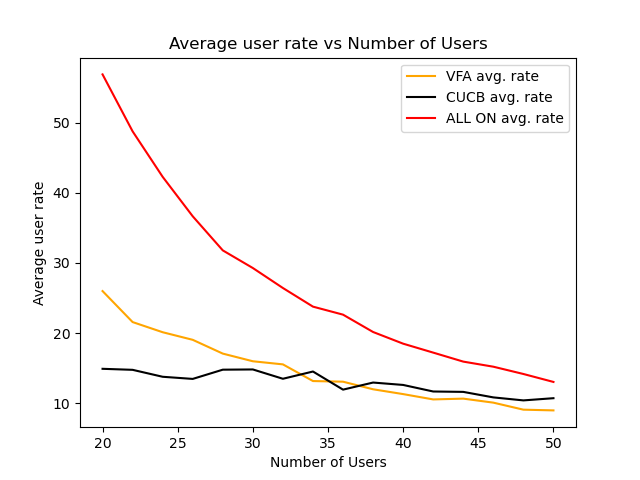}
\label{fig: Rate_vs_UE_Num}}
\hfil
\subfloat[]
{\includegraphics[width=0.32\textwidth]{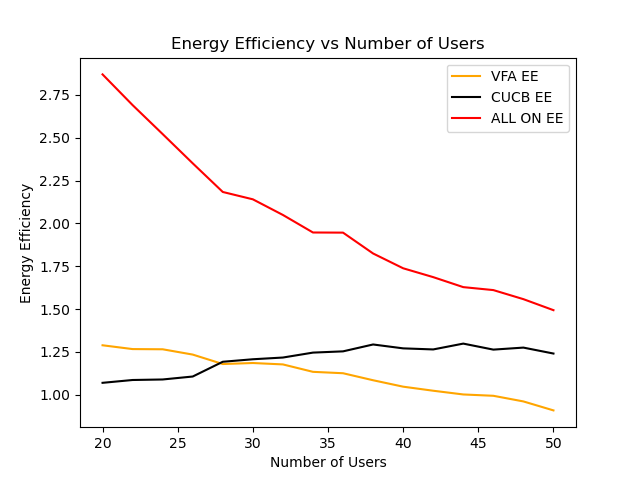}
\label{fig: EE_vs_UE_Num}}
\hfil
\subfloat[]
{\includegraphics[width=0.32\textwidth]{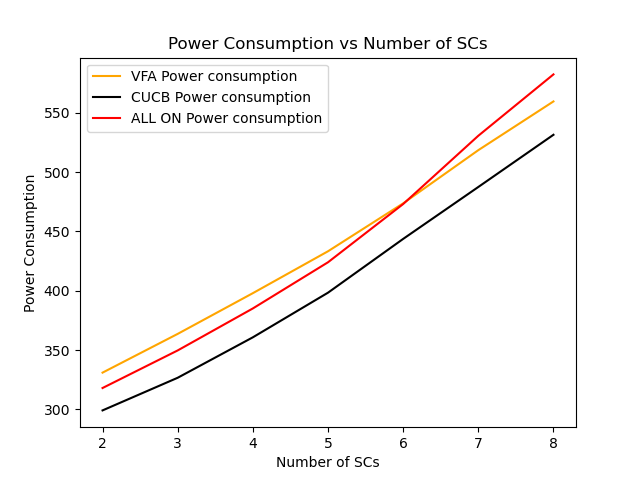}
\label{fig: PC_vs_Num_SCs}}
\hfil
\subfloat[]
{\includegraphics[width=0.32\textwidth]{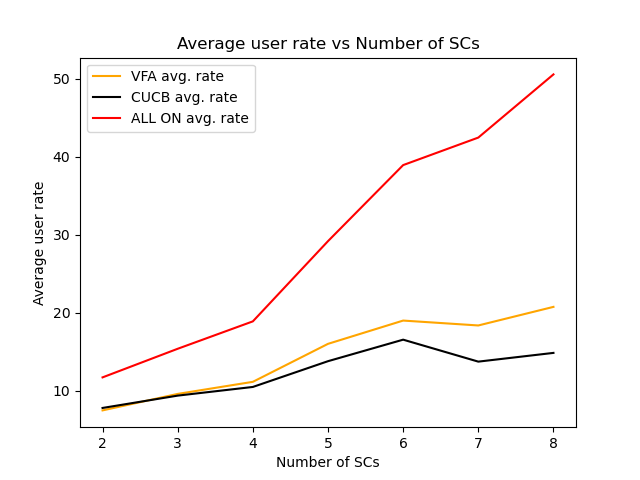}
\label{fig: Rate_vs_Num_SCs}}
\hfil
\subfloat[]
{\includegraphics[width=0.32\textwidth]{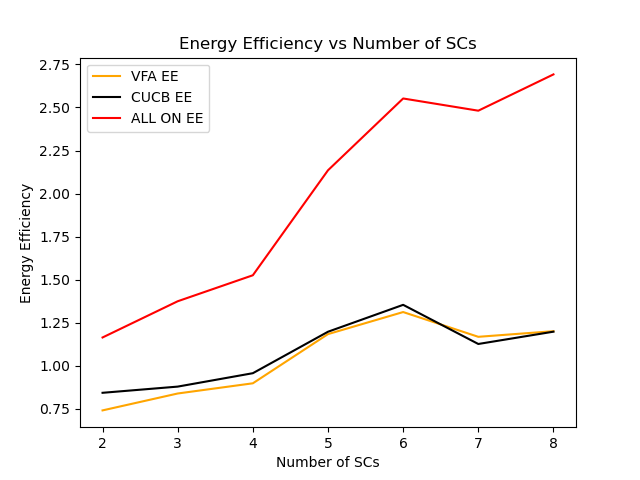}
\label{fig: EE_vs_Num_SCs}}
\caption{(a) Power Consumption, (b) Average \ac{UE} Data Rate, and (c) Energy Efficiency variation with the number of \acp{UE} for 5 \acp{SC}. (d) Power Consumption, (e) Average \ac{UE} Data Rate, and (f) Energy Efficiency variation with the number of \acp{SC} with 30 \acp{UE}.}
\label{fig:main_fig} 
\end{figure*}

\section{Numerical Results}
Our simulation framework consists of a \ac{MC} coverage disk in which \acp{SC} and \acp{UE} are deployed uniformly. The network parameters for the simulations are provided in Table I, while Table II presents the \ac{PC} characteristics of the various types of \acp{BS}. We have used only micro \acp{SC} for the simulations, however the results are similar for other types of \acp{SC} as well. The performance of the proposed mechanism is compared against the \texttt{ALL-ON} strategy (where no \ac{SC}-switching is implemented) and the \ac{SC} sleep strategy proposed in \cite{ozturk2021energy}, which we refer to as \texttt{VFA-SARSA}. The \texttt{VFA-SARSA} approach leverages the \ac{CDSA} architecture, and the \ac{SC} switching problem is formulated as an energy minimization problem subject to QoS constraints. As the state–action space grows exponentially with the number of \acp{SC}, an \ac{RL} algorithm called SARSA (state-action-reward-state-action), with Value Function Approximation (VFA), is employed to find the optimal ON/OFF pattern. Also, we consider that there is no \ac{CRE} parameter optimization done in either of these methods.
\begin{table}[t]
\centering
\caption{Simulation Parameters}
\begin{tabular}{|l|l|}
\hline
\textbf{Parameter} & \textbf{Value} \\
\hline
Macro cell radius & 500 m  \\
\hline
\ac{MC}/\ac{SC} Carrier frequency & 2.0 GHz  \\
                  
\hline
\ac{MC}/\ac{SC} Bandwidth  & 20 MHz \\

\hline
Pathloss Exponent ($\alpha_{p}$) & 4 \\
\hline
Noise Power Density & 3.98 x $10^{-21}$ W/Hz \\
\hline
$\mathcal{P}_1$, $\mathcal{P}_2$ & 100 (Both) \\
\hline
{$(\alpha,\beta)$-Approx. Oracle}
& $\alpha$ = 0.989, $\beta$ = 0.98 \\
\hline
\end{tabular}
\end{table}
\begin{table}[t]
\centering
\caption{Power Profiles for BSs \cite{auer2011much}}
\begin{tabular}{|l|c|c|c|c|}
    \hline
    \textbf{BS Type}
      & \textbf{Efficiency} $\eta$& \multicolumn{3}{c|}{\textbf{Power Consumption [W]}} \\
    \hline
    & 
      & \textbf{$P_S$} 
      & \textbf{$P_{oi}$}
      & \textbf{$P_{\rm{sleep}}$}  \\
    \hline
Macro  & 4.7 & 20   & 130   & 75 \\
\hline
RRH    & 2.8 & 20   & 84    & 56 \\
\hline
Micro  & 2.6 & 6.3  & 56    & 39 \\
\hline
Pico   & 4.0 & 0.13 & 6.8   & 4.3 \\
\hline
Femto  & 8.0 & 0.05 & 4.8   & 2.9 \\
\hline
\end{tabular}
\end{table} 
Fig.~\ref{fig: PC_vs_UE_Num} shows that the \ac{PC} of all the strategies increases with the number of \acp{UE}, as the load on the \acp{BS} increases. The \texttt{VFA-SARSA} performs better than \texttt{ALL-ON} until 27 \acp{UE}, then it starts to consume more power than \texttt{ALL-ON} due to the lack of appropriate load-balancing. Indeed, the increase in the load-dependent \ac{PC} is not compensated by the power savings due to switching off some \acp{SC}. The \ac{CUCB} outperforms the others by employing efficient load-balancing optimization after a \ac{SC} ON/OFF decision. As the number of \acp{UE} increases, \ac{CUCB} strategy imitates \texttt{ALL-ON} but still consumes less power.
Fig.~\ref{fig: Rate_vs_UE_Num} shows that the \texttt{ALL-ON} provides a higher data rate initially than the other two strategies due to no \acp{SC} switching. For low number of \acp{UE}, \texttt{VFA-SARSA} also provides better data rates as compared to \ac{CUCB}. But as soon as the number of \acp{UE} increases, \ac{CUCB} outperforms the \texttt{VFA-SARSA} because it can manage more \acp{UE} by optimizing the load of the network efficiently. As the number of \acp{UE} increases, both \ac{CUCB} and \texttt{VFA-SARSA} start to follow the \texttt{ALL-ON}.
As \ac{CUCB} outperforms the \texttt{VFA-SARSA} both in terms of data rate and \ac{PC} for dense \acp{UE}, its \ac{EE} increases with the number of \acp{UE}, as shown in Fig.~\ref{fig: EE_vs_UE_Num} . For a higher number of \acp{SC} in the network, a similar trend in performance is observed.
From the above analysis, we can conclude that the \ac{CUCB} is able to manage a larger number of \acp{UE} than the \texttt{VFA-SARSA} and \texttt{ALL-ON}, hence improving the capacity of the network without consuming more power.

Fig.~\ref{fig: PC_vs_Num_SCs} shows that the network \ac{PC} increases as the number of \acp{SC} deployed increases for all the three strategies. The \ac{CUCB} consumes less power than the others, while as \texttt{VFA-SARSA} initially (for lower number of \acp{SC}) consumes more power than the \texttt{ALL-ON} due to higher load. This is resolved after 6 \acp{SC} in the \ac{MC} coverage area. The higher energy savings comes at the cost of a lower data rate with respect to an increase in densification, as shown in Fig.~\ref{fig: Rate_vs_Num_SCs}. Nevertheless, this satisfies the \ac{QoS} requirement of 1 MBps. The \ac{CUCB} minimizes \ac{PC} by shutting down low-load \acp{SC} that increases the load of other \acp{SC} that are ON, and the \ac{MC}. This leads to lower data rates provided to \acp{UE}. Although the same holds for \texttt{VFA-SARSA}, it provides slightly better average \ac{UE} data rates. \texttt{ALL-ON} provides a better average \ac{UE} rate of all the strategies because all the \acp{SC} are ON with low loads, indicating that the impact of interference here is lower than the benefit of having more base stations. 
Fig.~\ref{fig: EE_vs_Num_SCs} shows the trends of the \acp{EE} versus the number of \acp{SC}. \texttt{ALL-ON} performs better than other strategies, as it provides a better average \ac{UE} data rate. \ac{EE} of \ac{CUCB} strategy is better than \texttt{VFA-SARSA} for lower number of \acp{SC} because it provides almost the same average \ac{UE} rate and consumes lower power, as can be seen from Figs.~\ref{fig: PC_vs_Num_SCs} and ~\ref{fig: Rate_vs_Num_SCs}. Then, as the number of \acp{SC} increase, both \ac{CUCB} and \texttt{VFA-SARSA} achieve the same \acp{EE}. For a larger number of \acp{UE} in the network, \ac{CUCB} performs better than \texttt{VFA-SARSA} both in terms of data rate as well as energy savings.

The performance for non-uniform \ac{UE} traffic distributions will be investigated in future work. 

\section{System Design Insights}
\begin{itemize}
    \item {\bf Clustered activation policy:} The results show that optimal SC sleep patterns are linked to spatio-temporal traffic patterns. Grouping \acp{SC} with correlated load profiles and applying the \ac{CUCB} policy can significantly reduce switching overhead while preserving performance.
    \item {\bf Exploration budget selection:} The observed sublinear regret behavior indicates that a limited exploration phase, preferably during low-traffic periods is sufficient for convergence, enabling stable long-term operation with minimal performance loss.
    \item {\bf Joint CRE and sleep optimization:} Powell-based CRE tuning substantially improves macro offloading efficiency during SC sleep. Coordinating CRE adjustments with SC activation leads to better load balancing and sustained user QoS compared to static CRE settings.
    \item {\bf Robustness to traffic variability:} This method adapts effectively to time-varying traffic, indicating that it can be deployed without relying on precise traffic forecasts, using only online measurements for policy updates.
\end{itemize}
\section{Conclusion}
This paper presented a bandit framework for switching ON/OFF \acp{SC} and efficiently optimizing the load of an ultra-dense heterogeneous network in order to reduce the \ac{PC} during low traffic periods. This mechanism employs a CMAB algorithm called CUCB to learn the best ON/OFF configuration of NR \acp{SC}. The decision outputs can be integrated into ESM functions in the RIC. A derivative-free continuous optimization procedure is then executed to tune CRE parameters in line with 3GPP RRM specifications, thereby optimizing the overall load distribution in the network. This not only saves energy but also increases the load-handling capability of the network. We compared this mechanism with the \texttt{ALL-ON} strategy and a state-of-the-art technique we refer to as \texttt{VFA-SARSA}. Simulation results show that the proposed \ac{CMAB} solution consumes less power than the \texttt{ALL-ON} strategy as well as the \texttt{VFA-SARSA} strategy while guaranteeing the required \ac{QoS} of \acp{UE}. Also, it was shown that the proposed method was able to handle a large number of \acp{UE} as compared to \texttt{VFA-SARSA}, hence increasing the capacity of the network.

\bibliographystyle{IEEEtran}
\bibliography{references}

\end{document}